\documentclass[11pt]{article}
\usepackage{moriond,epsfig}

\def\be{\begin{equation}}
\def\ee{\end{equation}}
\def\bea{\begin{eqnarray}}
\def\eea{\end{eqnarray}}

\newcounter{fig}
\newcommand{\mylist}[2]
  {\begin{list}{#1}{
      \setlength{\topsep}{1 pt}
      \setlength{\parsep}{0.0in}
      \setlength{\itemsep}{#2}
  }}
\newcommand{\newlist}[3]
  {\begin{list}{#1}{
      \setlength{\topsep}{0 pt}
      \setlength{\parsep}{0.0in}
      \setlength{\itemsep}{#2}
      \setlength{\leftmargin}{#3}
  }}
\newcommand{\newenum}[1]
  {
  \begin{list}{\arabic{fig}.}{\usecounter{fig}
      \setlength{\topsep}{0 pt}
      \setlength{\parsep}{0.0in}
      \setlength{\itemsep}{#1}
  }}

\newcommand{\newnum}
  {\begin{enumerate}{
      \setlength{\itemsep}{1 pt}
      \setlength{\parskip}{0 pt}
      \setlength{\parsep}{0 pt}
      \setlength{\topsep}{-20 pt}
      \setlength{\partopsep}{-20 pt}
  }}

\newcommand{\CP}{\mbox{\it CP}}
\newcommand{\ra}{\mbox{~$\rightarrow$}~}

\newcommand{\X}{\mbox{\it X}}

\newcommand{\Xz}{\mbox{\it X$^{\circ}$}}
\newcommand{\lm}{\mbox{\it l$^-$}}
\newcommand{\lp}{\mbox{\it l$^+$}}

\newcommand{\elp}{\mbox{$e^+$}}
\newcommand{\elm}{\mbox{$e^-$}}

\newcommand{\mum}{\mbox{$\mu^-$}}
\newcommand{\mup}{\mbox{$\mu^+$}}

\newcommand{\ggam}{\mbox{$\gamma$}}

\newcommand{\piz}{\mbox{$\pi^{\circ}$}}
\newcommand{\gpi}{\mbox{$\pi$}}
\newcommand{\pip}{\mbox{$\pi^+$}}
\newcommand{\pim}{\mbox{$\pi^-$}}

\newcommand{\Kp}{\mbox{$K^+$}}

\newcommand{\Kl}{\mbox{$K_{L}$}}

\newcommand{\gp}{\mbox{$p$}}
\newcommand{\agp}{\mbox{$\overline{p}$}}

\newcommand{\gL}{\mbox{$\Lambda$}}
\newcommand{\agL}{\mbox{$\overline{\Lambda}$}}

\newcommand{\gXi}{\mbox{$\Xi$}}

\newcommand{\Xim}{\mbox{$\Xi^-$}}
\newcommand{\aXim}{\mbox{$\overline{\Xi}$$^+$}}

\newcommand{\Sigp}{\mbox{$\Sigma^+$}}

\newcommand{\Omm}{\mbox{$\Omega^-$}}

\newcommand{\aOmm}{\mbox{$\overline{\Omega}$$^+$}}
\newcommand{\alphaLYt}%
{\makebox{$\alpha = {2\mbox{Re}(S^{\ast}P)}/(|S|^2 + |P|^2)$}}

\def\kp3pi{K^+ \rightarrow \pi^+ \pi^+ \pi^-}
\def\km3pi{K^- \rightarrow \pi^- \pi^- \pi^+}
\def\kpm3pi{K^{\pm} \rightarrow \pi^{\pm} \pi^+ \pi^-}

\begin{document}
\vspace*{4cm}
\title{Hints of New Physics in the Decay \boldmath{\Sigp\ra\gp\mup\mum}}

\author{E.\ Craig Dukes\footnote{Representing the HyperCP collaboration:
        Y.C.~Chen, 
        {\em (Academia Sinica)};
        W.S.~Choong, G.~Gidal, Y.~Fu, T.~Jones, K.B.~Luk, P.~Zyla,
        {\em (Berkeley and LBNL)};
        C.~James, J.~Volk,
        {\em (FNAL)};
        J.~Felix,
        {\em (Guanajuato)};
        R.A.~Burnstein, A.~Chakravorty, D.M.~Kaplan, L.M.~Lederman,
        W.~Luebke, D.~Rajaram, H.A.~Rubin, C.G.~White, S.L.~White,
        {\em (IIT, Chicago)};
        N.~Leros, J.P.~Perroud,
        {\em (Lausanne)};
        H.R.~Gustafson, M.J.~Longo, F.~Lopez, H.K.~Park,
        {\em (Michigan)};
        K.~Clark, C.M.~Jenkins,
        {\em (S.\ Alabama)};
        E.C.~Dukes, C.~Durandet, T.~Holmstrom, M.~Huang, L.C.~Lu, K.S.~Nelson,
        {\em (Virginia)}.
}}

\address{Physics Department, University of Virginia, \\
Charlottesville, VA USA}

\maketitle\abstracts{
The HyperCP (E871) experiment collected ${\sim}10^{9}$ hyperon
decays in the 1997 and 1999 Fermilab fixed-target running periods.
Using the data from the 1999 run, we report on the observation of
three isolated events with reconstructed masses consistent with the
hypothesis \Sigp\ra\gp\mup\mum.  This is the rarest baryon decay
ever observed.  The dimuon mass distribution is unexpectedly narrow,
suggesting the decay may proceed via an intermediate state of mass
$214.3{\pm}0.5$\,MeV/$c^2$.  This state is consistent with
a short-lived pseudoscalar sgoldstino with parity-conserving interactions
decaying into two unlike-sign muons.
}

\section{Introduction}

The decay \Sigp\ra\gp\lp\lm\ ($l = e, \mu$) is of interest because in the
standard model it is highly suppressed, with flavor-changing-neutral-current 
and weak-radiative leading diagrams.  Hence
observation of such a decay at a level greater than expected would
signal new physics.  Current experimental limits on such decays are
relatively weak:  there is an upper bound of $7{\times}10^{-6}$
on the \Sigp\ra\gp\elp\elm\ decay mode \cite{ang},
but no limit on the \Sigp\ra\gp\mup\mum\ decay mode.   Using 
the largest hyperon data sample ever collected, 
the HyperCP collaboration has searched for  \Sigp\ra\gp\mup\mum\
with hitherto unprecedented sensitivity.

\section{The HyperCP Apparatus}

\begin{figure}[htb]
\centerline{\includegraphics[width=4.25in]{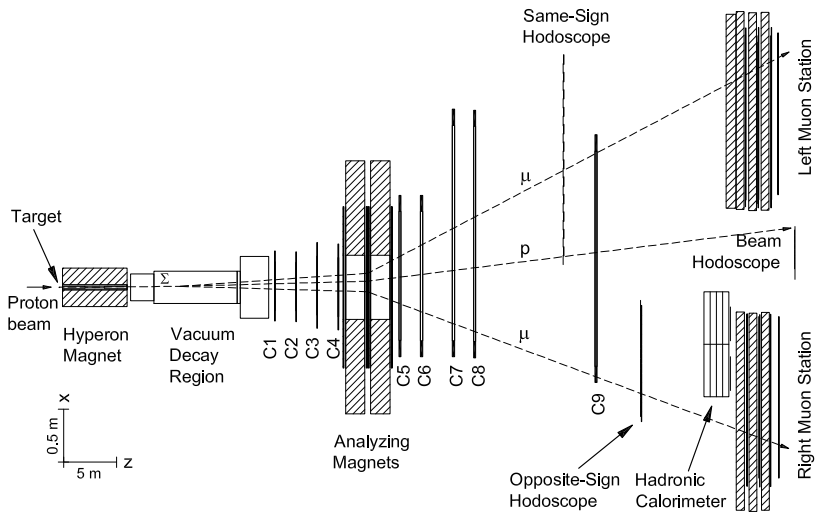}}
\caption{Plan view of the HyperCP spectrometer.}
\label{fig:spect_plan}
\end{figure}

The HyperCP experiment (Fig.~\ref{fig:spect_plan}) was designed 
primarily to investigate \CP\ violation in charged \gXi\ and \gL\ 
hyperon decays \cite{hypercp_nim}.  A charged-secondary beam of 
170\,GeV/$c$ average momentum
was produced by steering an 800\,GeV/$c$ proton beam onto a 
$2{\times}2$~mm$^2$ cross section Cu target followed by a curved channel
embedded in a 6~m long dipole magnet (hyperon magnet).
Charged particles were momentum analyzed in a magnetic 
spectrometer employing high-rate, narrow-pitch wire chambers.
The polarities of the hyperon and spectrometer magnets were periodically 
flipped to select oppositely charged secondary beams: the analysis 
reported here is from the positive-polarity data sample.
At the rear of the spectrometer were two identical muon stations 
positioned on either side of the secondary beam.
Each station had three sets of horizontal and vertical
proportional-tube planes, each behind 0.81~m of iron absorber,
as well as horizontal and vertical hodoscopes, which were used to 
trigger on muons.
The analysis reported herein is from 2.14 billion unlike-sign muon
triggers.

\section{Selecting Events}

\begin{figure}[htb]
\centering
\begin{minipage}[t]{3.3in}
\centerline{\includegraphics[height=3.0in]{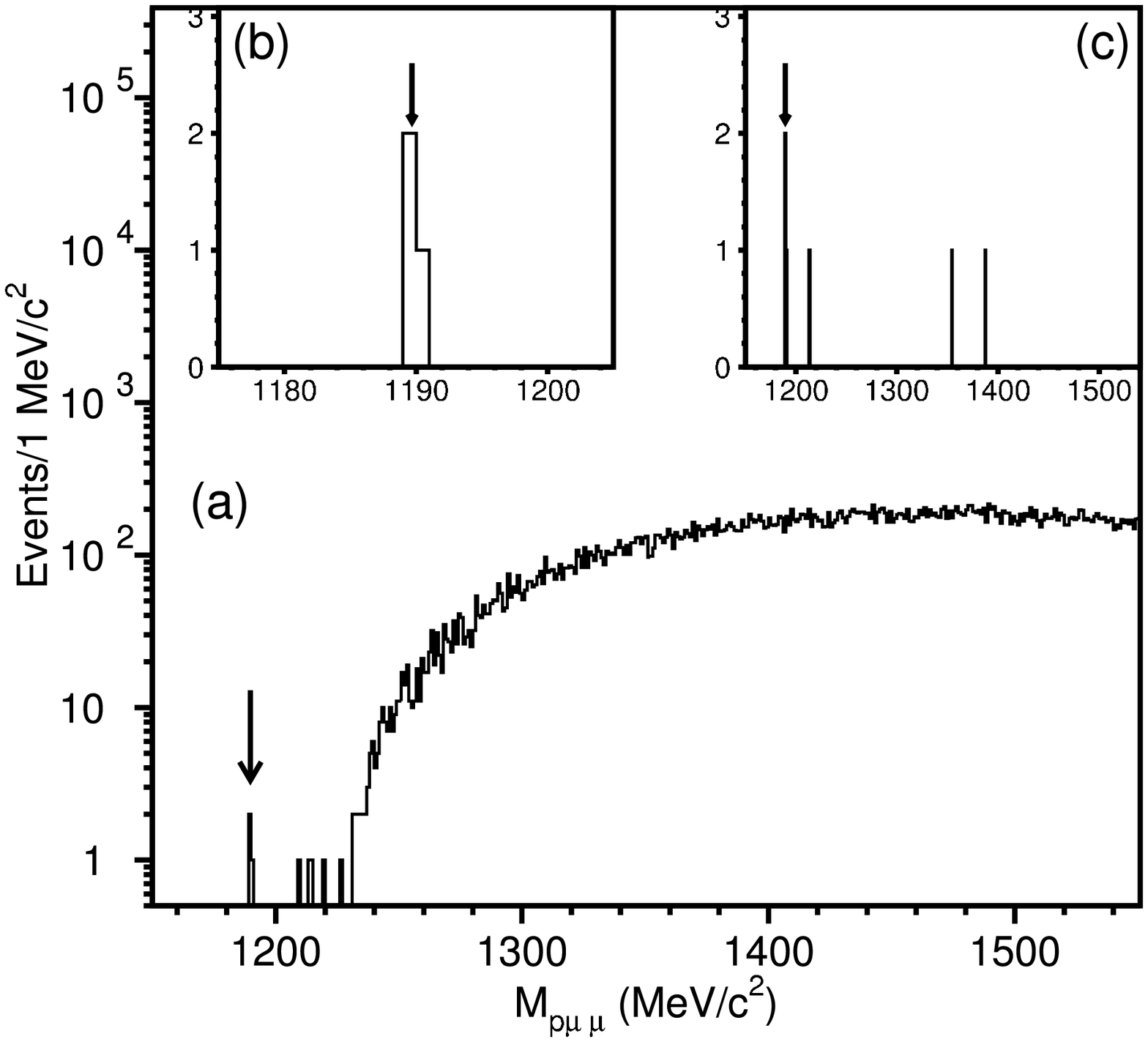}}
\caption{The \gp\mup\mum\ invariant mass: (a) after basic event-selection 
cuts; (b) in the \Sigp\ mass region; and (c) after
the application of the kaon-removal cuts.\hfill \mbox{}}
\label{fig:pmumu_mass}
\end{minipage}
\hfill
\begin{minipage}[t]{2.6in}
\centerline{\includegraphics[height=3.0in]{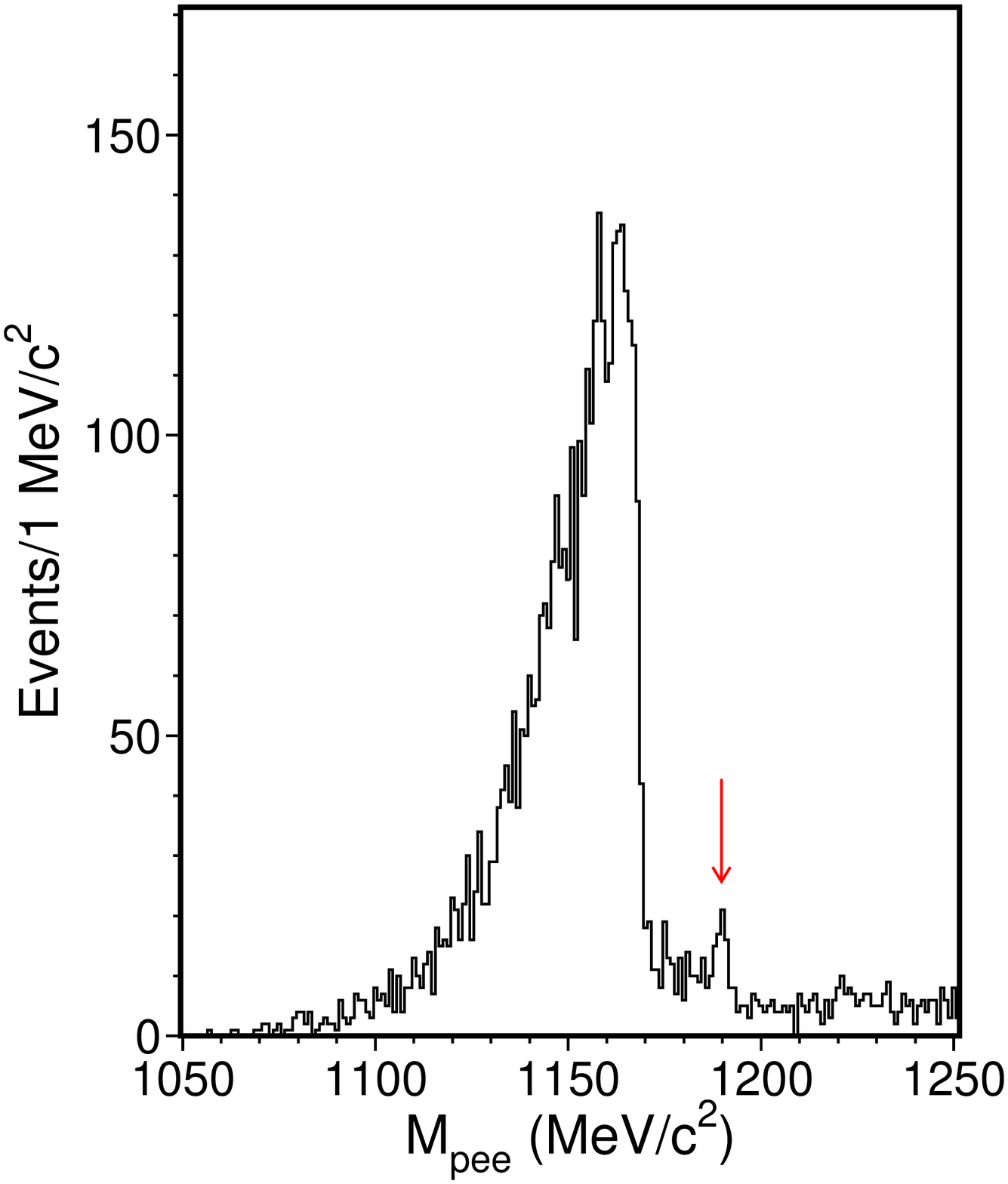}}
\caption{The \gp\elp\elm\ invariant mass from a preliminary
analysis.  A clear peak at the \Sigp\ mass is evident.
The fractions of events at the \Sigp\ mass coming from
true \Sigp\ra\gp\elp\elm\ decays and from \Sigp\ra\gp\ggam\ decays
with the \ggam\ converting to \elp\elm\ pairs, is not known.\hfill \mbox{}}
\label{fig:ee_mode}
\end{minipage}
\end{figure}

The basic event-selection cuts required:
(1) that two positively charged and one
    negatively charged track emanate from a common vertex with
    the distance-of-closest-approach less than 2.5~mm and the vertex-fit
    $\chi^2 < 1.5$;
(2) that the decay vertex lie well within the vacuum decay region;
(3) that the extrapolated track of the \Sigp\ point back to
    within 3.5 mm of the center of the target; 
(4) that there be two oppositely charged tracks each with 
    hits in two of the three muon proportional tube planes;
and
(5) that the highest momentum track not be a muon candidate
    and that it have the same sign charge as the secondary beam.
To eliminate kaon decays --- particularly \Kp\ra\pip\pim\pip\ 
and \Kp\ra\pip\mup\mum\ decays --- further cuts were made
on: (1) the ratio of the momentum of the non-muon track to the total 
momentum of the putative \Sigp: \begin{equation}
   f_{\rm hadron} = \frac{\textrm{``hadron'' momentum}}
                         {{\rm total}\ \Sigp\ {\rm momentum}}
   \geq 0.68,
\end{equation}
and (2) events with a \pip\mup\mum\ invariant mass within $3\sigma$
of the \Kp\ mass.
These two cuts eliminated essentially all of the kaon decays that
passed the basic event-selection cuts.  Figure \ref{fig:pmumu_mass}
shows the \gp\mup\mum\ invariant mass after application of the
basic event selection cuts.  Three events are found within
$1\,\sigma$ of the \Sigp\ mass, and about $20\,\sigma$ from the 
large kaon-decay background, which with the application of the 
kaon-removal cuts is reduced to only four events.

Much effort was spent in investigating whether the three signal events
were indeed real \Sigp\ decays.  Backgrounds from other hyperon decays 
were negligible.  There is no other positively charged hyperon, 
so hyperon background events would have to come from
anti-hyperon decays, such as \aXim\ra\agL\pip\ra\agp\pip\pip\ or 
\aOmm\ra\agL\Kp\ra\agp\pip\Kp, where the pion or kaon and the 
anti-proton are somehow misidentified as 
muons.  In addition, the highest momentum track in such decays --- almost
always the antiproton --- has the wrong
sign charge and hence would not pass the $f_{\rm hadron}$ cut.
Hence the likelihood of the three signal events being anti-hyperons
is negligible.

More plausible hyperon-decay backgrounds are the dominant \Sigp\ decay, 
\Sigp\ra\gp\piz\ra\gp\ggam\ggam\ (BR = 0.53), or the
weak-radiative decay \Sigp\ra\gp\ggam\ (BR = $1.2{\times}10^{-3}$), 
with the gamma converting to a unlike-sign muon pair.  However, the 
probability of a gamma converting to a \mup\mum\ pair in the 
windows of the vacuum decay region, at $\sim 10^{-7}$, is negligibly small.  
A search for the \Sigp\ra\gp\elp\elm\ decay mode shows evidence 
of a signal at the \Sigp\ mass, as can be seen in Fig.~\ref{fig:ee_mode}.  
If gamma conversions were indeed the source of the unlike-sign muons, 
then since the conversion rate
\ggam\ra\elp\elm\ is $\mathcal{O}(10^5)$ greater than \ggam\ra\mup\mum,
one would expect far more events in the \Sigp\ra\gp\elp\elm\ mode:
that is clearly not the case.  Note that the analysis of the 
\Sigp\ra\gp\elp\elm\ decay mode is much more difficult than
the \Sigp\ra\gp\mup\mum\ mode since HyperCP has no electron identification,
and \Sigp\ra\gp\elp\elm\ events from \Sigp\ra\gp\ggam\ conversions
are a non-negligible background.

We have analyzed the HyperCP negative-polarity data sample,
which is about half the size of the positive-polarity sample,
using the same cuts, and, as expected, we find no events at the
\Sigp\ mass satisfying the 
event-selection cuts.  We also searched the single-muon trigger 
sample --- some thirty times larger than the dimuon trigger sample 
(and prescaled by a factor of ten) --- and, again, as expected,
we find no events below 1200 MeV/$c^2$, indicating that backgrounds 
do not appear to survive the event-selection cuts.  Nor did 
relaxing the event-selection cuts add any background events in
the \Sigp\ mass region.

\section{Determining the Branching Ratio}

Since the spectrometer acceptance was not perfect, the form factors
had to be known for the branching ratio to be extracted.  The four
form factors describing the \Sigp\ra\gp\mup\mum\ decay cannot be 
calculated ab initio, 
but were extracted from the branching ratio and asymmetry of the 
\Sigp\ra\gp\ggam\ decay and limits on the \Sigp\ra\gp\elp\elm\ branching 
ratio.  The decay used as the normalization mode in determining the 
branching ratio was \Sigp\ra\gp\piz, where one of the gammas from the 
\piz\ra\ggam\ggam\ decay converted to an \elp\elm\ pair, and the
other was not detected: HyperCP had no gamma detectors.  We find:
\begin{equation}
   \mathcal{B}(\Sigp\ra\gp\mup\mum) = 
      [8.6^{+6.6}_{-5.4}{\rm (stat)}{\pm}5.5{\rm (syst)}]{\times}10^{-8}.
\end{equation}
Using a uniform phase-space, rather than form-factor, model for the 
\Sigp\ra\gp\mup\mum\ decay increases the branching ratio by about 50\%.
If we assume that the three signal events are all from some unknown
background then we obtain an upper limit at 90\ C.L. of
$\mathcal{B}(\Sigp\ra\gp\mup\mum) < 3.4{\times}10^{-7}$.

A theoretical calculation by Bergstr\"{o}m et al.\ estimates the 
branching ratio to be $\sim 10^{-8}$ \cite{Bergstrom}.  A more recent 
calculation by He at al.\ predicts
the branching ratio to lie between $1.6{\times}10^{-8}$ and 
$9.0{\times}10^{-8}$ \cite{Heprd}.

\section{The Dimuon Mass Distribution}

\begin{figure}
\centering
\begin{minipage}[t]{4.0in}
\centerline{\includegraphics[width=4.0in]{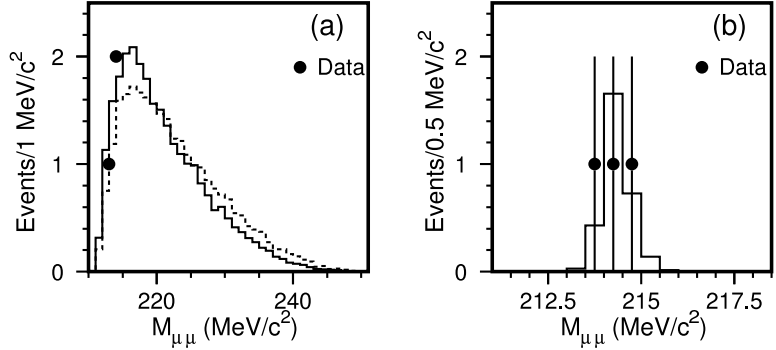}}
\caption{The \mup\mum\ invariant mass of the three signal events with 
superimposed (a) Monte Carlo form factor decays (solid histogram) and uniform 
phase-space decays (dashed histogram), and (b) Monte Carlo 
$\Sigp\ra\gp X^0$, $X^0\ra\mup\mum$ events generated with
$m_{X^0} = 214.3 {\rm MeV}/c^2$.\hfill \mbox{}}
\label{fig:mass_mumu}
\end{minipage}
\end{figure}

Unexpectedly, as shown in Fig.~\ref{fig:mass_mumu}, all three signal 
events have dimuon masses within 1 MeV/$c^2$ of each 
other, which is the mass resolution of the HyperCP spectrometer.
The probability that
the three masses would all have the same value anywhere in the
allowed kinematic range is about 1\%.  Varrying the form factors
within their allowed ranges does not increase this probability
significantly.  As pointed out by He et al.\ \cite{Heplb},
it is highly unlikely that this narrow mass distribution could be due
to the formation of a muonium-bound state, despite the fact that the 
mean dimuon mass is only 3 MeV/$c^2$ higher than twice the muon mass.
This suggests that the
decay proceeds via an hitherto unknown intermediate state $X^0$ of 
mass $214.3{\pm}0.5$~MeV/$c^2$ with a branching ratio 
$\mathcal{B}(\Sigp\ra\gp\X^0, X^0\ra\mup\mum) =
[3.1^{+2.4}_{-1.9}{\rm (stat)}{\pm}1.5{\rm (syst)}]{\times}10^{-8}$.

Unfortunately, HyperCP can say little about
the putative $X^0$ particle's properties, outside of its mass.
Searches in the kaon sector eliminate the possibility that
it is a parity violating; HyperCP, for example, does not find
any evidence of a dimuon mass peak in their \Kp\ra\pim\mup\mum\ data sample.
If $X^0$ were a 
vector particle then KTeV would have seen evidence of its
decay in their \Kl\ra\ggam\mup\mum\ data sample \cite{KTeV}.
Hence, assuming its properties are not too exotic,
one must assume that $X^0$ is a parity conserving
pseudoscalar or axial vector.

It has been pointed out by Gorbunov and Rubakov that such a particle
would be consistent with the sgoldstino, the superpartner to
the goldstino \cite{Gorbunov06}.  The sgoldstino is expected to be spin 0, 
all of its other properties are ill determined:  it can be light, 
long or short lived, there should be two, a scalar and a pseudoscalar, 
and its interactions can be flavor conserving and flavor violating, 
and may or may not conserve parity.  The branching ratio to dimuons 
can be large.

\section{Conclusions}

This observation begs to be confirmed or refuted.  Unfortunately, 
HyperCP has exhausted their available data, 
and with the Tevatron fixed-target 
program over at Fermilab, there are no prospects for further
running.  The only other similar such hyperon decay that is 
kinematically allowed is \Omm\ra\Xim\mup\mum.  Although HyperCP 
has the world's largest \Omm\ sample, with an expected branching 
ratio of $\mathcal{O}(10^{-6})$ \cite{Heplb,deshpande}, at best one would 
find only a few events.
However, four-body kaon decay limits are comparatively weak,
and modes such as \Kl\ra\gpi\gpi\Xz\ are expected to have
branching ratios $\mathcal{O}(10^{-8})$ \cite{Heplb,Gorbunov06,deshpande}
and should be pursued, as well as the other possible channels discussed, 
for example, in Ref.~\cite{Gorbunov06}.

We finally note that these results have recently been published
in Physical Review Letters \cite{hypercp}.

\section*{Acknowledgments}

We thank the organizers for a most interesting and stimulating
conference.  We are indebted to the staffs of Fermilab and
the participating institutions for their vital contributions.
This work was supported by the U.S. Department of Energy and
the National Science Council of Taiwan, R.O.C\@.  E.C.D. and
K.S.N. were partially supported by the Institute for Nuclear
and Particle Physics at the University of Virginia.  K.B.L. was 
partially supported by the Miller Institute.

\end{document}